\documentclass{article}
\usepackage{pdfpages}
\usepackage{xcolor}
\usepackage[colorlinks=true, allcolors=blue]{hyperref}
\usepackage{graphicx}
\usepackage{csquotes}
\usepackage{cite}
\usepackage{caption}
\usepackage{subcaption}
\usepackage{caption}
\usepackage{authblk}
\usepackage{amsmath}
\usepackage[top=5cm, bottom=5cm, left=3cm, right=2cm, headsep=1.5cm, heightrounded=true]{geometry}
\usepackage{fancyhdr}
\graphicspath{ {{images/}} }

\title{ Model-Independent Reconstruction of $f(T)$ Gravity Using Genetic Algorithms }
\author{Redouane El Ouardi$^{1}$\thanks{ \texttt{redouane.elouardi.d23@ump.ac.ma}}}
\author{Amine Bouali$^{1,2,3}$\thanks{ \texttt{a1.bouali@ump.ac.ma}}}
\author{Imad El Bojaddaini$^{1}$\thanks{ \texttt{i.elbojaddaini@ump.ac.ma}}}
\author{Ahmed Errahmani$^{1,2}$\thanks{ \texttt{ahmederrahmani1@yahoo.fr}}}
\author{Taoufik Ouali$^{1,2}$\thanks{ \texttt{t.ouali@ump.ac.ma}}}
\affil {\textit{$^{1}$Laboratory of Physics of Matter and Radiation, }\\
		\centering \textit{ Mohammed I University, BP 717,  Morocco}\\
		\textit{$^{2}$Astrophysical and Cosmological Center, Faculty of Sciences, BP 717, Morocco}\\
		\textit{$^{3}$Higher School of Education and Training, Mohammed I University, BP 717, Morocco}}

\begin{document}
	\maketitle

	In this paper, we use genetic algorithms, a specific machine learning technique, to achieve a model-independent reconstruction of $f(T)$ gravity. By using $H(z)$ data derived from cosmic chronometers and radial Baryon Acoustic Oscillation method, including the latest Dark Energy Spectroscopic Instrument (DESI) data,  we reconstruct the Hubble rate which is  the basis parameter for reconstructing $f(T)$ gravity  without any assumptions. In this reconstruction process, we use the current value of the Hubble rate, $H_0$, derived by genetic algorithms.  The reconstructed $f(T)$ function is consistent with the standard $\Lambda$CDM cosmology within the 1$\sigma$ confidence level across a broad temporal range. The mean  $f(T)$ curve, adopting a quadratic form, prompts us to parametrize it using a second degree polynomial. This quadratic deviation from the $\Lambda$CDM scenario is mildly favored by the data.

	% Keywords
	\newpage
	% Introduction
	\section{Introduction}\

    The acceleration of the Universe's expansion \cite{SupernovaSearchTeam:1998fmf,SupernovaCosmologyProject:1998vns}, observed in the late of 1990s, represents a major challenge in modern physics. This discovery led to the hypothesis of an exotic form of energy, known as dark energy, within the framework of general relativity. The $\Lambda$CDM model, which incorporates this energy in the form of a cosmological constant and cold dark matter, is currently the reference model in cosmology, as it successfully explains a wide range of observations, from galaxies to large-scale structures of the Universe.  However, the recent accumulation of high-precision cosmological data, such as observations of type Ia supernovae, cosmic microwave background, and baryon acoustic oscillations, has revealed tensions and inconsistencies within the $\Lambda$CDM model, suggesting the possibility of missing or incomplete features.	\
	
	One of the most striking examples of these tensions is the discrepancy at high and low redshift in the Hubble constant, $H_{0}$, which measures the current expansion rate of the Universe. Observations of the cosmic microwave background (CMB) suggest a value of $H_0$ $=$ 67.40$ \hspace*{0.1 cm}\pm \hspace*{0.1 cm} 0.50$ km s$^{-1}$ Mpc$^{-1}$ based
	on the  $\Lambda$CDM paradigm \cite{Wong:2019kwg, Planck:2018vyg}, while the SH0ES team recently published a measurement of $H_{0}$, based on the direct distance scale, of  $H_0 = 73.03$$\hspace*{0.1 cm} \pm \hspace*{0.1 cm}1.04$ \text{km s$^{-1}$Mpc$^{-1}$}  \cite{Riess:2021jrx}. This tension, reaching almost 5$\sigma$, reveals a significant statistical inconsistency and raises questions about the validity of the $\Lambda$CDM model.  Additionally, there is another tension related to $\sigma_8$, the parameter that measures the amplitude of matter density fluctuations on a scale of 8 $h^{-1}$ Mpc, where $h$ is the Hubble constant in units of 100 \text{km s$^{-1}$Mpc$^{-1}$}. This tension arises between the values obtained from the CMB measurements based on the $\Lambda$CDM model and those from galaxy surveys and weak gravitational lensing observations at low redshifts \cite{Poulin:2022sgp, Dahmani:2023bsb}. These tensions and other issues lead physicists to explore the potential physical causes of these phenomena \cite{Abdalla:2022yfr}.\
	
	In order to resolve these challenges, various modifications have been proposed, affecting both early and late cosmology. Among these modifications, we cite models of dynamic dark energy \cite{Dahmani:2023bsb, mhamdi2023comparing, Karwal:2016vyq, Vagnozzi:2021gjh, bouhmadi2018more, errahmani2006high, enkhili2024diagnostic, belkacemi2020interacting}, the introduction of new relativistic species in the early Universe \cite{Gelmini:2019deq}, and theories of modified gravity \cite{Errahmani:2024ran, mhamdi2024cosmological, bouali2023observational, Koussour:2022dcb,harko2022palatini, Clifton:2011jh, bargach2021dynamical, ElOuardi}.\

	Among the various theories of modified gravity, three main approaches can be distinguished. The first one consists in extending the Einstein-Hilbert action, the pillar of general relativity, by introducing arbitrary functions of geometric quantities based on curvature, such as $ f(R) $ gravity \cite{DeFelice:2010aj} and $f(G)$ gravity \cite{Nojiri:2005jg, ladghami20234d}. The second approach relies on the teleparallel equivalent of general relativity (TEGR) \cite{ Maluf:2013gaa, Aldrovandi:2013wha, Krssak:2018ywd} and explores modifications based on torsion, such as $f(T)$ gravity \cite{Bengochea:2008gz, Linder:2010py}, $f(T, T_{G})$ gravity \cite{Kofinas:2014owa,  Capozziello:2016eaz} ($T_{G}$ represents the teleparallel analogue of the Gauss-Bonnet term), etc. The third approach known as symmetric teleparallel gravity uses non-metricity scalar $Q$ and introduces a function $f(Q)$ in the Lagrangian to develop new extended theories of gravity \cite{Heisenberg:2023lru, mhamdi2024cosmological, enkhili2024cosmological, koussour2023cosmic, mhamdi2024observational}. 
	
	Within the framework of torsion-based gravity theories, $f(T)$ gravity provides a novel alternative for explaining the accelerated expansion of the Universe without the need for dark energy \cite{Bengochea:2008gz}, as well as an alternative to inflation without requiring an inflaton field \cite{Ferraro:2006jd, Ferraro:2008ey}. However, a key challenge for any modified gravity theory is selecting a viable function from the numerous possible options. The general process involves considering various specific forms within a general class, applying them in a cosmological framework, predicting dynamic behaviors at both the background and perturbation levels, and using observational data to constrain the involved parameters or exclude certain theoretical forms \cite{Cai:2019bdh}. The literature often
	studies three viable $f(T)$ gravity scenarios: the power law, the exponential square root form, and the exponential form \cite{Nesseris:2013jea, Bengochea:2008gz}.\

	In this paper, we proceed differently by using machine learning to construct $f(T)$ gravity. Machine learning (ML) has achieved numerous successes in cosmology \cite{moriwaki2023machine0}, particularly in distinguishing between standard and modified gravity theories using statistically similar weak lensing maps \cite{Peel:2018aei} and in reducing scatter in cluster mass estimates \cite{Ho:2019zap}. Among the most promising ML approaches are genetic algorithms. Genetic algorithm (GA) is a computer simulation that stochastically searches for an optimal solution to a given problem. It evaluates a set of candidate solutions, called "individuals," grouped into a "population." Through an iterative process of crossover and mutation, guided by a "fitness" function, individuals are selected based on how well they solve the problem. For example, parts of two or more functions (individuals) can be combined to form new functions by crossover processes, while mutations make random changes, such as altering coefficients or exponents. This process of evolution allows genetic algorithms to find the best solutions to complex problems effectively. Over generations, the population evolves and converges toward an optimal solution.\
	
	Genetic algorithms are robust and promising methods that efficiently explore the search space to find optimal solutions. They have been successfully used in various fields, such as particle physics \cite{Akrami:2009hp, Allanach:2004my}, cosmology \cite{Nesseris:2010ep,  Nesseris:2013bia, Alestas:2022ml}, astronomy, and astrophysics \cite{rajpaul2012genetic, Ho:2019zap}, among others.\
	
	In this paper, we explore a new approach to reconstruct the form of the function $f(T)$ within the framework of $f(T)$ cosmology, using genetic algorithms. Previous studies have used Gaussian processes (GP) to reconstruct $f(T)$ and $f(R, T)$ gravity \cite{Cai:2019bdh, Ren:2022aeo, Ren:2021tfi, fortunato2024search}. However, GA offers a distinct advantage over GP as they do not require any prior assumptions about the underlying physical model \cite{Arjona:2020kco}. On the other hand, GP requires the choice of a kernel function. In this regard, one study employed GA to select the appropriate kernel function that provides the best results \cite{Bernardo}. GA efficiently explores the search space to find a solution close to the global optimum rather than a local one. This is the first application of genetic algorithms to reconstruct the forms of modified gravities. Our objective is to determine the form of $f(T)$ gravity in a model-independent manner, relying solely on the Hubble data. To achieve this, we use the  $H(z)$ data obtained from distinct methods:  the differential age method \cite{Jimenez:2001gg} and the radial BAO method \cite{gaztanaga2009clustering}, including the latest Dark Energy Spectroscopic Instrument (DESI) BAO \cite{adame2024desi}. $f(T)$ cosmology presents a crucial advantage for this approach, because the torsion scalar $T$ is directly related to $H(z)$ by the relation $T = -6H^2$. Through GA, we reconstruct $H(z)$ from the observational data. This reconstruction allows us to determine the form of $f(T)$ without any assumptions.\

	The structure of this article is as follows: In Section \ref{section2}, we provide an overview of the theoretical framework for $f(T)$ gravity along with its cosmological equations. Section \ref{section3} details the methodology behind genetic algorithms and their application to derive the Hubble function $H(z)$ from observational data. Following this, in Section \ref{section4}, we utilize the reconstructed $H(z)$ to establish the shape of the $f(T)$ function independently of any model. Section \ref{section5} summarizes the key findings and draws the conclusions.

	\section{$f(T)$ gravity and cosmology}\label{section2}
	
	\hspace{0.6 cm}In this section, we briefly introduce $f(T)$ gravity and its applications within a cosmological framework. The torsion-based formulations of gravity utilize tetrad fields $e^A {}_\mu $, which serve as dynamic variables. The Greek indices represent spacetime coordinates and Latin indices denote tangent space. These tetrads form an orthonormal basis for the tangent space at each point on the manifold. The relationship between the metric $g_{\mu\nu}$ and the tetrads is expressed through the equation
	
	\begin{equation}
		g_{\mu\nu} = \eta_{AB} e^A{}_{\mu} e^B{}_{\nu},
	\end{equation}
	where $\eta_{AB}$ is the tangent space metric, generally considered the Minkowski metric with the signature $\eta_{AB} = \text{diag}(1, -1, -1, -1)$.
	
	In teleparallel gravity, the torsionless Levi-Civita connection is replaced by the Weitzenböck connection, defined as $W^\lambda{}_{\nu\mu} \equiv e_A{}^\lambda \partial_\mu e^A{}_\nu$ \cite{t}. This connection leads to zero curvature but non-zero torsion. The associated torsion tensor can then be expressed as
	
	\begin{equation}
		T^\lambda {}_{\mu \nu} \equiv W^\lambda{}_{\nu\mu} - W^\lambda{}_{\mu\nu} = e_{A}{}^\lambda (\partial_\mu e^A{}_\nu - \partial_\nu e^A{}_\mu).
	\end{equation}
	
	The Lagrangian for the teleparallel equivalent of general relativity, based on the torsion scalar, $T$, is formulated through contractions of this torsion tensor as
	
	\begin{equation}\label{z}
		T \equiv \frac{1}{4} T^{\rho\mu\nu} T_{\rho\mu\nu} + \frac{1}{2} T^{\rho\mu\nu} T_{\nu\mu\rho} -  T_{\rho\mu}^{\ \ \rho} T^{\nu\mu}_{\ \ \ \nu}.
	\end{equation}

	Inspired by $f(R)$ extension of general relativity, we begin with the teleparallel equivalent of general relativity to develop various gravitational modifications. Specifically, the simplest extension involves generalizing the torsion scalar $T$ in the action as ($T + f(T)$) gravity formulation. The action for this theory is given by \cite{Cai:2019bdh}
	\begin{equation} \label{eq}
		S = \frac{1}{16\pi G} \int d^4 x \hspace{0.1 cm}e \left[ T + f(T)\right] + \int d^4 x \hspace{0.1 cm}e\hspace{0.1 cm} L_m ,
	\end{equation}
	where $G$ is the gravitational constant, $e$ $=$ det$(e^A{}_\mu$) $= \sqrt{-g}$ represents the determinant of the tetrad field, and $L_m$ is the matter Lagrangian included to account for the presence of matter.\\
	
	By varying the action, Eq. (\ref{eq}), with respect to the tetrads, we obtain the following field equations as \cite{Cai:2019bdh}	
	\begin{equation*}
		e^{-1}\partial_{\mu}(e e^\rho{}_{A}S_{\rho}{}^{\mu\nu})[1+f_T]+e^\rho{}_{A}S_{\rho}{}^{\mu\nu}\partial_{\mu}(T)f_{TT}\\
		-[1+f_T]e_A{}^{\lambda}T^{\rho}{}_{\mu\lambda}S_{\rho}{}^{\nu\mu}+\frac{1}{4}e_A^{\ \ \nu}[T+f(T)]
	\end{equation*}
	\begin{equation} \label{q}
		= 4\pi G e^\rho{}_{A}\overset{\hspace{-0.4 cm}\textbf{em}}{T_\rho{}^{\nu}},
	\end{equation}
	where $f_T = \partial f /\partial T$, $f_{TT} = \partial^{2} f /\partial T^{2}$, with $\overset{\hspace{-0.4 cm}\textbf{em}}{T_\rho{}^{ \nu}}$ denoting the matter\footnote{namely dark and baryonic matter.} energy-momentum tensor, and $S_{\rho}{}^{\mu\nu}$ is the super-potential defined as
	\begin{equation}
		S_\rho{}^{\phantom{\rho}\mu\nu} \equiv \frac{1}{2} \left( K^{\mu\nu}{}_{\ \rho }+ \delta^\mu_\rho T^{\alpha\nu}{}_{\ \alpha} - \delta^\nu_\rho T^{\alpha\mu}{}_{\ \alpha} \right),
	\end{equation}
	with $K^{\mu\nu}{}_{\rho } \equiv -\frac{1}{2} \left( T^{\mu\nu}{}_{\ \ \rho } - T^{\nu\mu}{}_{\ \rho} - T_{\rho}^{\ \ \mu\nu} \right) $ is the contorsion tensor.\\

	To apply $f(T)$ gravity within a cosmological framework, we consider a homogeneous and isotropic flat geometry, described by the tetrad $e^A{}_\mu = \text{diag}(1, a(t),  a(t), a(t))$. This particular tetrad setup corresponds to the spatially flat Friedmann-Lemaître-Robertson-Walker (FLRW) metric
	
	\begin{equation}
		ds^2 = dt^2 - a^2(t) \left[ dr^2 + r^2 (d\theta^2 + sin^2 \theta d\phi^2) \right],
	\end{equation}
	where $a(t)$ the scale factor.
	
	Inserting the vierbein $e^A {}_\mu$ into the field equations, Eq. (\ref{q}), we obtain the Friedmann equations  for
	$f(T)$ cosmology as
	
	\begin{equation} \label{8}
		H^2 = \frac{8\pi G}{3} \rho_m - \frac{f}{6} + \frac{Tf_T}{3},
	\end{equation}	
	\begin{equation} \label{9}	 
		\dot{H} = -\frac{4\pi G(\rho_m + P_m)}{1 + f_T + 2Tf_{TT}},
	\end{equation}
	where $H \equiv \frac{\dot{a}}{a}$ represents the Hubble parameter, and the dot above a variable denotes the derivative with respect to cosmic time $t$. Furthermore, $\rho_m$ and $P_m$ denote the energy density and pressure of matter, respectively. The expression of the torsion scalar in the FRW geometry writes
	\begin{equation}\label{10}
		T = - 6 H^2.
	\end{equation}
	From the first Friedmann equation, Eq. (\ref{8}), it is possible to define an effective dark energy sector of gravitational origin within $f(T)$ cosmology. The energy density and pressure of this sector can be respectively defined as follows
	
	\begin{equation}\label{11}
		\rho_{DE} \equiv \frac{3}{8\pi G} \left[ -\frac{f}{6} + \frac{Tf_T}{3} \right], \\
	\end{equation}
	\begin{equation} \label{12}
		P_{DE} \equiv \frac{1}{16\pi G} \left[ \frac{f - f_T T + 2T^2 f_{TT}}{1 + f_T + 2Tf_{TT}} \right],
	\end{equation}
	and the corresponding equation-of-state parameter can then be expressed as follows
	\begin{equation}\label{13}
		w \equiv \frac{P_{DE}}{\rho_{DE}} = -\frac{f/T - f_T + 2Tf_{TT}}{[1 + f_T + 2Tf_{TT}][f/T - 2f_T]}.
	\end{equation}
	Finally, the system of cosmological equations is closed by the inclusion of the matter conservation equation 
	\begin{equation}
		\dot{\rho}_m + 3H(\rho_m + P_m) = 0,
	\end{equation}
	thus, from Eqs. (\ref{8}) and (\ref{9}), the conservation of the effective energy is written as
	\begin{equation}
		\dot{\rho}_{DE} + 3H(\rho_{DE} + P_{DE}) = 0.
	\end{equation}\\

	\section{Reconstruction of the Hubble Parameter}\label{section3}
	\hspace{0.6cm}In this section, we first present the process of genetic algorithms, and then we apply it to reconstruct a model from observational data.

\begin{table}[!t]
\centering
\caption{The $H(z)$ dataset and corresponding uncertainties, $\sigma_H$, employed in our analysis (expressed in km s$^{-1}$Mpc$^{-1}$).}
\label{table1}
\begin{tabular}{|c|c|c|c|c|c|c|c|}
\hline
\multicolumn{8}{|c|}{CC data} \\
\hline
\textbf{$z$} & \textbf{$H(z)$} & \textbf{$\sigma_{H}$} & Ref. & \textbf{$z$} & \textbf{$H(z)$} & \textbf{$\sigma_{H}$} & Ref. \\
\hline
0.07  & 69.0  & 19.6  & \cite{zhang2014four1} &  0.4783 & 83.8  & 10.2 & \cite{Moresco2016mzx}\\
0.09  & 69.0  & 12.0  & \cite{jimenez2003constraints}  & 0.48  & 97.0  & 62.0  & \cite{stern2010cosmic1} \\
0.12  & 68.6  & 26.2  & \cite{zhang2014four1} &  0.5929 & 107.0 & 15.5 & \cite{moresco2012improved} \\
0.17  & 83.0  & 8.0   & \cite{stern2010cosmic}   & 0.6797 & 95.0  & 10.5 &  \cite{moresco2012improved} \\
0.1791 & 78.0  & 6.2  & \cite{moresco2012improved}& 0.75 &  98.8 & 33.6 & \cite{borghi2022toward}\\
0.1993 & 78.0  & 6.9  & \cite{moresco2012improved} & 0.7812 & 96.5  & 12.5 &  \cite{moresco2012improved}\\
0.2  & 72.9  & 29.6  & \cite{zhang2014four1}  & 0.8754 & 124.5 & 17.4 &  \cite{moresco2012improved} \\
0.27  & 77.0  & 14.0  & \cite{stern2010cosmic}  & 0.88  & 90.0  & 40.0  & \cite{stern2010cosmic1}   \\
0.28  & 88.8  & 36.6  & \cite{zhang2014four1}  & 0.90  & 117.0 & 23.0  & \cite{stern2010cosmic}  \\
0.3519 & 85.5  & 15.7 & \cite{moresco2012improved} &  1.037  & 133.5 & 17.6 &\cite{moresco2012improved}\\
0.3802 & 86.2  & 14.6 & \cite{Moresco2016mzx} & 1.30  & 168.0 & 17.0  & \cite{stern2010cosmic} \\
0.4  & 95.0  & 17.0  & \cite{stern2010cosmic}  & 1.363  & 160.0 & 33.8 & \cite{Moresco2015cya} \\
0.4004 & 79.9  & 11.4 & \cite{Moresco2016mzx} & 1.43  & 177.0 & 18.0  & \cite{stern2010cosmic}  \\
0.4247 & 90.4  & 12.8 & \cite{Moresco2016mzx} & 1.53  & 140.0 & 14.0  & \cite{stern2010cosmic}\\
0.4497 & 96.3  & 14.4 & \cite{Moresco2016mzx} & 1.75  & 202.0 & 40.0  & \cite{stern2010cosmic}\\
0.47  & 89.0  & 49.6  & \cite{ratsimbazafy2017age} & 1.965  & 186.5 & 50.6 & \cite{Moresco2015cya}  \\
\hline
\multicolumn{8}{|c|}{BAO data} \\
\hline
\textbf{$z$} & \textbf{$H(z)$} & \textbf{$\sigma_{H}$} & Ref. & \textbf{$z$} & \textbf{$H(z)$} & \textbf{$\sigma_{H}$} & Ref. \\
\hline
0.24 & 79.69 & 2.99 & \cite{gaztanaga2009clustering}  & 0.57 &96.80 & 3.40  & \cite{anderson2014clustering}   \\
0.30 & 81.70 & 6.22  & \cite{oka2014simultaneous} & 0.59 & 98.48 & 3.19  & \cite{wang2017clustering}  \\
0.31 & 78.17 & 6.74  & \cite{wang2017clustering}  & 0.6 & 87.90 & 6.10 & \cite{blake2012wigglez}  \\
0.34 & 83.17 & 6.74  & \cite{gaztanaga2009clustering} & 0.61 & 97.30 & 2.10  &  \cite{alam2017clustering} \\
0.35 & 88.1  & 9.45  & \cite{chuang2013modelling}   & 0.64 & 98.82 & 2.99  & \cite{wang2017clustering}  \\
0.36 & 79.93 & 3.93  & \cite{wang2017clustering}  & 0.978 &113.72 & 14.63  & \cite{zhao2019clustering}  \\
0.38 & 81.50 & 1.90  & \cite{alam2017clustering}  & 1.23 & 131.44 & 12.42  & \cite{zhao2019clustering}\\
0.40 & 82.04 & 2.03  & \cite{wang2017clustering}& 1.48 & 153.81 & 6.39 &  \cite{neveux2020completed}  \\
0.43 & 86.45 & 3.68  & \cite{gaztanaga2009clustering}   & 1.526 & 148.11 & 12.71 &  \cite{zhao2019clustering} \\
0.44 & 82.60 & 7.80  & \cite{blake2012wigglez}  & 1.944 & 172.63 & 14.79 &  \cite{zhao2019clustering}\\
0.44 & 84.81 & 1.83 &   \cite{wang2017clustering}& 2.3 & 224 & 8 & \cite{Busca_2013} \\
0.48 & 87.79 & 2.03 &  \cite{wang2017clustering} & 2.36 & 226.0 & 8&\cite{font2014quasar}  \\
0.56 & 93.33 & 2.32  &  \cite{wang2017clustering}  & 2.4 & 227.8 & 5.61 & \cite{des2017baryon}  \\
0.57 & 87.60 & 7.80  & \cite{chuang2013clustering}    & & & & \\
\hline 
\multicolumn{8}{|c|}{DESI data} \\
\hline
\textbf{$z$} & \textbf{$H(z)$} & \textbf{$\sigma_{H}$} & Ref. & \textbf{$z$} & \textbf{$H(z)$} & \textbf{$\sigma_{H}$} & Ref. \\
\hline
0.51 & 97.21 & 2.83  & \cite{adame2024desi}& 1.32 & 147.58 & 4.49  & \cite{adame2024desi}  \\
0.71 & 101.57 & 3.04  & \cite{adame2024desi} & 2.33 & 239.38 & 4.80  & \cite{adame2024desi}  \\
0.93 & 114.07 & 2.24  & \cite{adame2024desi} &&&& \\
\hline
\end{tabular}
\end{table}

	\subsection{Genetic Algorithms}
	
	\hspace{0.6 cm}Genetic algorithms operate by mimicking the notion of biological evolution through natural selection, where a set of candidate solutions, referred to as \enquote{individuals} (in our case, functions), is grouped into a \enquote{population} that evolves over time. The performance of each member is evaluated by a \enquote{fitness} function, which, in our analysis, is represented by a statistic chi-square, $\chi^2$, indicating how well each individual corresponds to the data. The individuals with the best fit in each generation are selected using a tournament selection method (see Ref. \cite{bogdanos2009genetic} for more details). Subsequently, two stochastic operations, mutation and crossover, are applied. Crossover involves combining segments of two or more functions (individuals) to create new functions, while mutations introduce random changes, such as modifying coefficients or exponents. This process is then repeated thousands of times to ensure convergence, allowing the population to evolve gradually toward an optimal solution.\\

Genetic algorithms are a machine learning technique specialized in unsupervised symbolic regression of data, where they analytically reconstruct a function that describes the data, using one or more variables. The algorithm begins with an initial population of candidate solutions, usually generated randomly using a predefined set of functions referred to as the \enquote{grammar} (e.g. polynomial, exponential, sine, logarithmic, etc.). This initial population represents the first generation of the algorithm, which evolves over multiple generations through stochastic operations such as crossover and mutation. To illustrate these operations, let us consider two example functions: $f_0(x) = 1 + 2x + 3x^2$ and $g_0(x) = \sin(x) + \exp(x)$. The mutation operation randomly modifies the terms in these functions. For instance, the functions might transform as follows: $f_0(x) \rightarrow f_1(x) = 1 + 2x + x^2$ and $g_0(x) \rightarrow g_1(x) = \sin(x^2) + \exp(x)$. In the first case, the coefficient of the third term changes from $3$ to $1$, while in the second case, the term $x$ is replaced by $x^2$. On the other hand, the crossover operation combines terms from two functions to create new ones. For example, combining $1 + 2x$ from $f_1(x)$ and $\sin(x^2)$ from $g_1(x)$ results in $f_1(x) \rightarrow f_2(x) = 1 + 2x + \sin(x^2)$. Similarly, merging the remaining terms produce $g_1(x) \rightarrow g_2(x) = x^2 + \exp(x)$. In each generation, the best-fitting functions are selected to proceed to the next generation. Note that the first generation does not affect the final solution, they only affect the rate of convergence of the GA. This process is repeated thousands of times with different random seeds in order to effectively explore the functional space to ensure convergence toward the optimal function that best fits the data. Convergence is reached when the $\chi^2$ value stabilizes, showing no improvement or change over a few hundred generations.
\\

	The final output of the GA code is a function that is both continuous and differentiable. However, GA does not inherently estimate the errors associated with the reconstructed functions, which is crucial for the statistical interpretation of the data. To address this limitation, it is necessary to employ a path integral method, as described in \cite{Nesseris_2012, Nesseris:2013bia}. This method estimates errors by analytically calculating a path integral over all possible functions that GA might explore. This approach to error reconstruction has been evaluated and reported in reference \cite{Nesseris_2012}. It has been shown to yield error estimates that are both smoother and comparable to those obtained through the bootstrap Monte Carlo method.

\subsection{Error estimates}\

The error estimates of the derived best-fit function are obtained through the implementation of the “path integral” method, initially proposed by the authors of Ref. \cite{Nesseris_2012}. The likelihood functional is

\begin{equation}
    \mathcal{L}=\mathcal{N} \exp \left(-\chi^2(f) / 2\right),
\end{equation}
where $f(x)$ represents the function reconstructed by the genetic algorithm, and $\chi^2(f)$ denotes the associated chi-squared for $N$ data points $(x_i, y_i, \sigma_i)$, defined as
\begin{equation}\label{17x}
    \chi^2(f) \equiv \sum_{i=1}^N\left(\frac{y_i-f\left(x_i\right)}{\sigma_i}\right)^2.
\end{equation}\
Determining the normalization constant $\mathcal{N}$ is considerably more complex than in the case of a normal distribution, as it requires integration over all possible functions $f(x)$ or in other words, performing a “path integral”
\begin{equation}\label{18x}
    \int \mathcal{D} f \mathcal{N} \exp \left(-\chi^2(f) / 2\right)=1.
\end{equation}
Here, $\mathcal{D}f$ denotes integration over all possible values of $f(x)$. This comprehensive integration is necessary because, during the operation of the GA, a wide variety of candidate functions are explored due to the stochastic nature of mutation and crossover operators. Although many of these candidate functions may not provide optimal fits and are eventually discarded, they still contribute to the overall likelihood. Consequently, their influence must be accounted for when calculating the error estimates of the best-fit.\

The infinitesimal measure, $\mathcal{D}f$, can be expressed as $\mathcal{D}f = \prod_{i=1}^N df_i$, where $df_i$ and $f_i$ denote $df(x_i)$ and $f(x_i)$, respectively. For the time being, we assume that the function values $f(x_i)$ and $f(x_j)$ at distinct points $x_i$ and $x_j$ are independent. Therefore, Eq. (\ref{18x}) can be reformulated as

\begin{align*}
\int \mathcal{D} f \mathcal{L} &= \int_{-\infty}^{+\infty} \prod_{i=1}^N d f_i \mathcal{N} \exp \left(-\frac{1}{2} \sum_{i=1}^N \left( \frac{y_i - f_i}{\sigma_i} \right)^2 \right) \\
&= \prod_{i=1}^N \int_{-\infty}^{+\infty} d f_i \mathcal{N} \exp \left( -\frac{1}{2} \left( \frac{y_i - f_i}{\sigma_i} \right)^2 \right) \\
&= \mathcal{N} \cdot (2 \pi)^{N/2} \prod_{i=1}^N \sigma_i \\
&= 1
\end{align*}
which leads to $\mathcal{N} = \left[(2\pi)^{N/2} \prod_{i=1}^{N} \sigma_i\right]^{-1}$. Consequently, the likelihood becomes
\begin{equation}
    \mathcal{L}=\frac{1}{(2 \pi)^{N / 2} \prod_{i=1}^N \sigma_i} \exp \left(-\chi^2(f) / 2\right).
\end{equation}
Considering our assumption that the function $f$ evaluated at each point $x_i$ is independent

\begin{equation}
    \mathcal{L}_i \equiv \frac{1}{(2 \pi)^{1 / 2} \sigma_i} \exp \left(-\frac{1}{2}\left(\frac{y_i-f_i}{\sigma_i}\right)^2\right).
\end{equation}\

Now, we can calculate the 1$\sigma$ error, $\delta f_i$, around the best-fit function $f_{\text{bf}}(x)$ at a point $x_i$ as
\begin{equation*}
C I\left(x_i, \delta f_i\right)  =\int_{f_{b f}\left(x_i\right)-\delta f_i}^{f_{b f}\left(x_i\right)+\delta f_i} d f_i \frac{1}{(2 \pi)^{1 / 2} \sigma_i} \exp \left(-\frac{1}{2}\left(\frac{y_i-f_i}{\sigma_i}\right)^2\right)
\end{equation*}
\begin{equation}\label{19x}
 =\frac{1}{2}\left(\operatorname{erf}\left(\frac{\delta \mathrm{f}_{\mathrm{i}}+\mathrm{f}_{\mathrm{bf}}\left(\mathrm{x}_{\mathrm{i}}\right)-\mathrm{y}_{\mathrm{i}}}{\sqrt{2} \sigma_{\mathrm{i}}}\right)+\operatorname{erf}\left(\frac{\delta \mathrm{f}_{\mathrm{i}}-\mathrm{f}_{\mathrm{bf}}\left(\mathrm{x}_{\mathrm{i}}\right)+\mathrm{y}_{\mathrm{i}}}{\sqrt{2} \sigma_{\mathrm{i}}}\right)\right).
\end{equation}\

%\color{black}
To determine the errors $\delta f_i$ corresponding to the $1\sigma$ error of a normal distribution, we numerically solve the following equation %derived from Eq. (3.13).
\begin{equation}
    C I\left(x_i, \delta f_i\right)=\operatorname{erf}(1 / \sqrt{2}),
\end{equation}
while it is possible to compute the $1\sigma$ error, $\delta f_i$, of the best-fit function $f_{\mathrm{bf}}(x)$ at each point $x_i$, this approach yields discrete error estimates at individual data points. Such pointwise estimates do not align with our objective of obtaining a smooth, continuous, and differentiable error function. For that, it is necessary to create a new chi-squared defined as
\begin{equation}\label{24x}
    \chi_{C I}^2\left(\delta f_i\right)=\sum_{i=1}^N\left(C I\left(x_i, \delta f_i\right)-\operatorname{erf}(1 / \sqrt{2})\right)^2.
\end{equation}
We also parameterize $\delta f$ using a second order polynomial form, $\delta f(x) = a + bx + cx^2$. The coefficients $(a, b, c)$ are determined by minimizing the combined chi-squared function, $\chi^2(f_{\mathrm{bf}} + \delta f) + \chi^2_{CI}(\delta f)$, where $\chi^2$ is defined in Eq. (\ref{17x}) and $\chi^2_{CI}$ in Eq. (\ref{24x}). This approach ensures that the $1\sigma$ confidence region for the best-fit function $f_{\mathrm{bf}}(x)$ lies within the interval $[f_{\mathrm{bf}}(x) - \delta f(x),\ f_{\mathrm{bf}}(x) + \delta f(x)]$. The reason why we perform a simultaneous fit of both chi-squares to ensure the resulting error region reflects only minor variations around the best-fit solution, rather than the entire infinite dimensional functional space explored by the genetic algorithm.\

The approach described above applies to uncorrelated data. In cases where data points are correlated, it is necessary to incorporate the chi-squared term accounting for these correlations. For more details on the path integral approach to the error estimation in the context of correlated data, we refer the reader to Ref. \cite{Nesseris_2012}. For instance, if the best-fit function derived by the genetic algorithm is the Hubble parameter, $H(z)$, this procedure can be adapted to determine the corresponding errors. Subsequently, the errors associated with each quantity related to $H(z)$ can be derived using error propagation techniques \cite{Arjona:2020kco}. For example, in the case of the deceleration parameter, we have
   \begin{equation}
     q_{\mathrm{GA}}(z)=-1+(1+z) \frac{d \ln H_{\mathrm{GA}}}{d z},  
   \end{equation} 
which implies that the error is
\begin{equation*}
\delta q(z)  =(1+z) \delta\left[\frac{d \ln H}{d z}\right] \end{equation*}
\begin{equation*}
=(1+z) \frac{d}{d z}[\delta \ln H] 
\end{equation*}
\begin{equation}
 =(1+z) \frac{d}{d z}\left[\frac{\delta H}{H_{\mathrm{GA}}}\right].
\end{equation}

	\subsection{Observational Hubble Data}\label{subsection3.3}\

    In our analysis, we utilize Hubble measurements obtained through two primary approaches. The first approach involves the clustering of galaxies or quasars, enabling a direct measurement of the Hubble expansion by identifying the Baryon Acoustic Oscillation (BAO) peak in the radial direction \cite{gaztanaga2009clustering}. The second approach relies on the differential age technique, commonly referred to as the cosmic chronometers (CC) method. This method is based on the relationship between the Hubble parameter and the time derivative of the redshift of distant objects, such as massive elliptical galaxies. The relationship is expressed as $H(z) = -\frac{1}{(1+z)}\frac{dz}{dt}$ \cite{Jimenez:2001gg}, which enables the estimation of $H(z)$ by measuring the relative ages of these objects at different redshifts.\\

In our work, we use 64 $H(z)$ data points of the Hubble parameter, which include 32 points from CC measurements and 32 points from BAO data as presented in \cite{Yang:2024kdo}. The BAO dataset consists of 27 points from previous BAO data combining independent datasets such as WiggleZ \cite{blake2012wigglez}, BOSS DR12 \cite{alam2017clustering}, and eBOSS DR16 \cite{gil2020completed, bautista2021completed, avila2020completed, hou2021completed, neveux2020completed, des2020completed}, along with five points from the most recent Dark Energy Spectroscopic Instrument (DESI) BAO results \cite{adame2024desi}. Within the CC dataset, 15 data points were provided by Moresco et al. \cite{moresco2012improved, Moresco2016mzx, Moresco2015cya}, all obtained using the same method. For more details on the correlation between these points, we refer the reader to reference \cite{moresco2020setting}. In our reconstruction, we take into account the covariance between these points, as detailed in their open-source program\footnote{\url{https://gitlab.com/mmoresco/CCcovariance/-/tree/master}}. This data is summarized in Table \ref{table1}.

To reconstruct the history of cosmic dynamics evolution from the $H(z)$ data, we perform a model-independent reconstruction of the Hubble parameter using genetic algorithms. For the likelihood $\chi^2$ of the $H(z)$ data, we use the following expressions, depending on whether the datasets are correlated or uncorrelated.

For correlated datasets, the chi-squared is given by
\begin{equation}
    \chi^2_{Corr} = \bigg( H_i - H_{\text{GA}}(z_i) \bigg)^T C_{ij}^{-1} \bigg( H_j - H_{\text{GA}}(z_j) \bigg),
\end{equation}
while for uncorrelated datasets, it is written as 
\begin{equation}
    \chi_{UnCorr}^2 = \sum_{i=1}^N \left( \frac{H_i - H_{\text{GA}}(z_i)}{\sigma_{H,i}} \right)^2,
\end{equation}
where $H_i$ and $\sigma_{H,i}$ represent the observed values of the Hubble parameter and their associated uncertainties at each redshift $z_i$, respectively. $H_{\text{GA}}(z)$ corresponds to the theoretical prediction of the Hubble parameter derived from genetic algorithms, while $C_{ij}^{-1}$ is the inverse covariance matrix of the correlated data. Finally, $N$ denotes the number of data points considered.\\

The total $\chi^2$ is then given by
\begin{equation}
    \chi^2_H = \chi^2_{Corr} + \chi^2_{UnCorr}.
\end{equation}

\begin{figure}[t]
		\centering
		\includegraphics[width=0.53\textwidth,height=0.22\textheight]{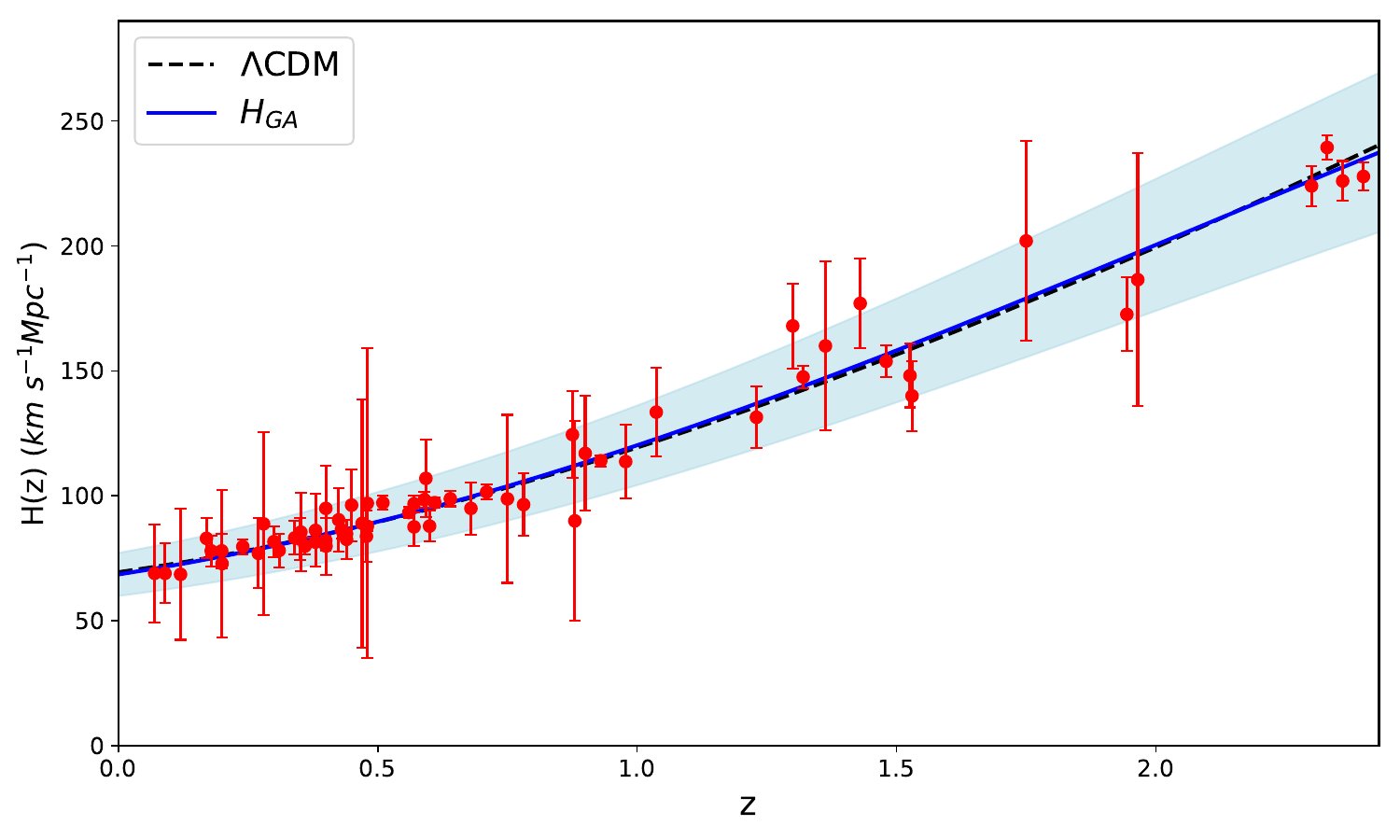}
        %,height=0.27\textheight]{Figures/Figure1.pdf} 
		\caption{The $H(z)$ data compilation, along with the $\Lambda$CDM best-fit (dashed black curve) and the GA best-fit (solid blue curve). The blue region represents the $1\sigma$ confidence level for the GA model, while the actual data points are depicted as red points.}
		\label{H} 
	\end{figure}

	It is important to note that when deriving a model from a dataset using genetic algorithms, we do not make any prior assumptions, such as a specific dark energy model or a flat Universe. This ensures that the obtained results do not depend on a particular choice. Furthermore, due to the very nature of how genetic algorithms operate, one can ensure that the final solution does not prematurely converge to a local optimum, but rather approaches the global one. For the dataset in Table \ref{table1}, the best solution $H_{GA}(z)$, derived from genetic algorithms and which minimizes $\chi_H^2$, is obtained as follows
	\begin{equation}
		H_{GA}(z) = H_{0} \hspace{0.1 cm} \Big(1 + z\hspace{0.1 cm}\big(0.682 + 0.222z - 0.036z^2\big)^2\Big),
	\end{equation}
	where $H_{0} = (68.56 \pm 8.59)$ km/s/Mpc. The best-fit $\chi^2$ value for the GA model is 41.43, compared to 42.47 for the $\Lambda$CDM best-fit model, with the best-fit values $\Omega_{m0} = 0.279 \pm 0.014$ and $H_0 = 69.45 \pm 1.03$ km/s/Mpc, obtained through an MCMC analysis. The GA model accurately fits the observed data, achieving a minimal value of $\chi^2$ comparable to that of the $\Lambda$CDM best-fit model. Fig. \ref{H} shows the reconstructed Hubble parameter compared to the $\Lambda$CDM model.

There are other non-parametric techniques commonly used for model-independent reconstructions of cosmological functions, such as Gaussian Processes (GP), Smoothing Methods (SM), and Artificial Neural Networks (ANN). In Ref. \cite{vazirnia}, the authors applied GP, SM, and GA to reconstruct the Hubble parameter and the luminosity distance, finding that the three methods yielded similar results. Among these techniques, GP and ANN are the most widely used in cosmology. However, GP reconstructions require the choice of both a kernel function and a fiducial model for the mean function, as previously mentioned. The choice of the kernel can sometimes influence the reconstruction outcome \cite{mukherjee}. Similarly, for ANN, the selection of hyperparameters, such as activation functions and the number of layers, can affect the reconstruction quality, and generally, a larger amount of training data is needed to ensure reliable results. An advantage of the approach based on GA is its independence from any predefined assumptions about the form of the reconstructed functions. Furthermore, GA naturally provides analytical functions that describe the provided data.

	\section{Reconstruction of $f(T)$ gravity}\label{section4}\
    
In the preceding subsection, we used genetic algorithms on the $H(z)$ data to achieve a model-independent reconstruction of $H(z)$. Thus, we can now use this reconstructed form of $H(z)$ to derive  the function $f(T)$ itself.

\begin{figure}[t]\label{figj} 
		\hspace{-0.58 cm}
		\raisebox{-0.33cm}{\includegraphics[width=0.575\textwidth]{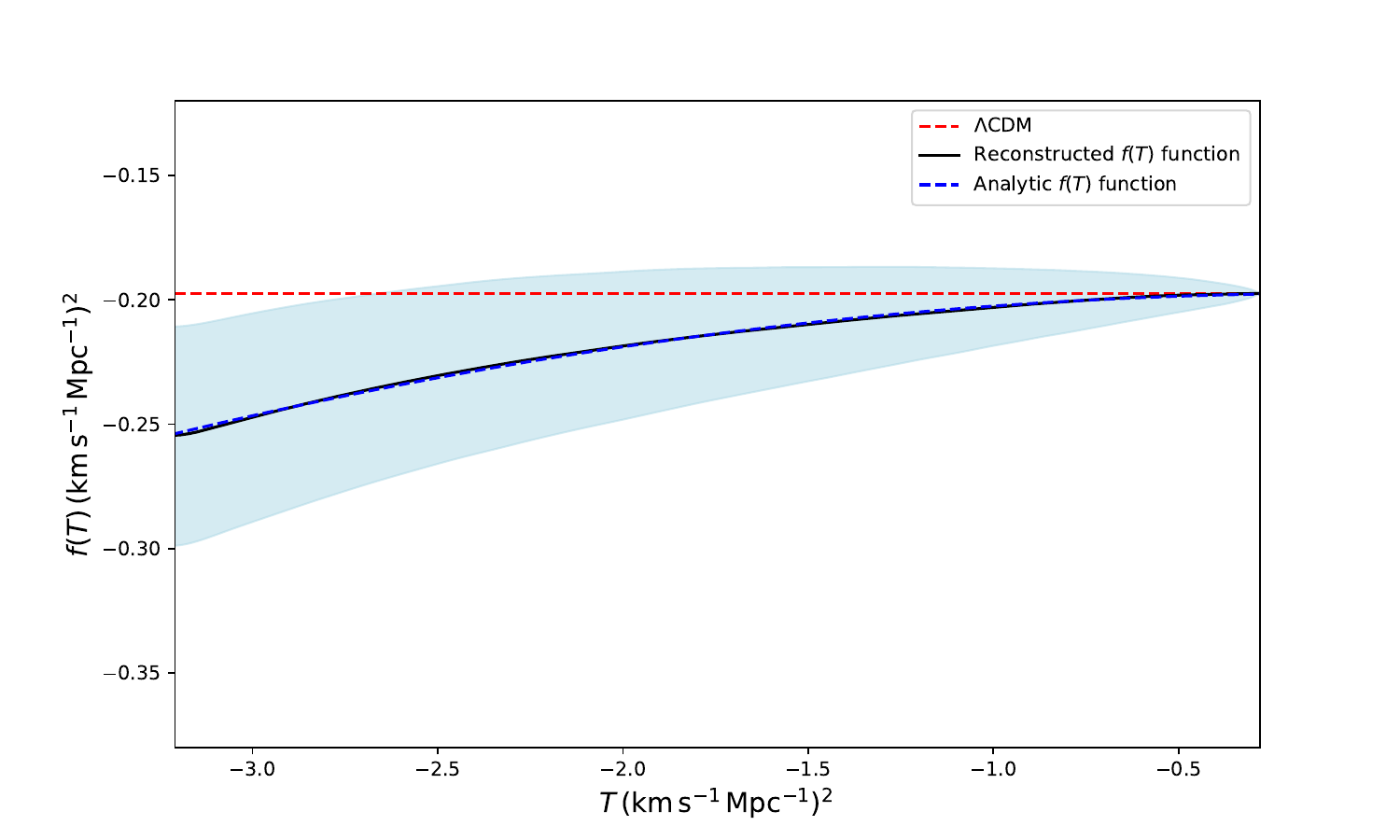}}
		\hspace{-0.8 cm}
		\raisebox{-0.2cm}{\includegraphics[width=0.5\textwidth]{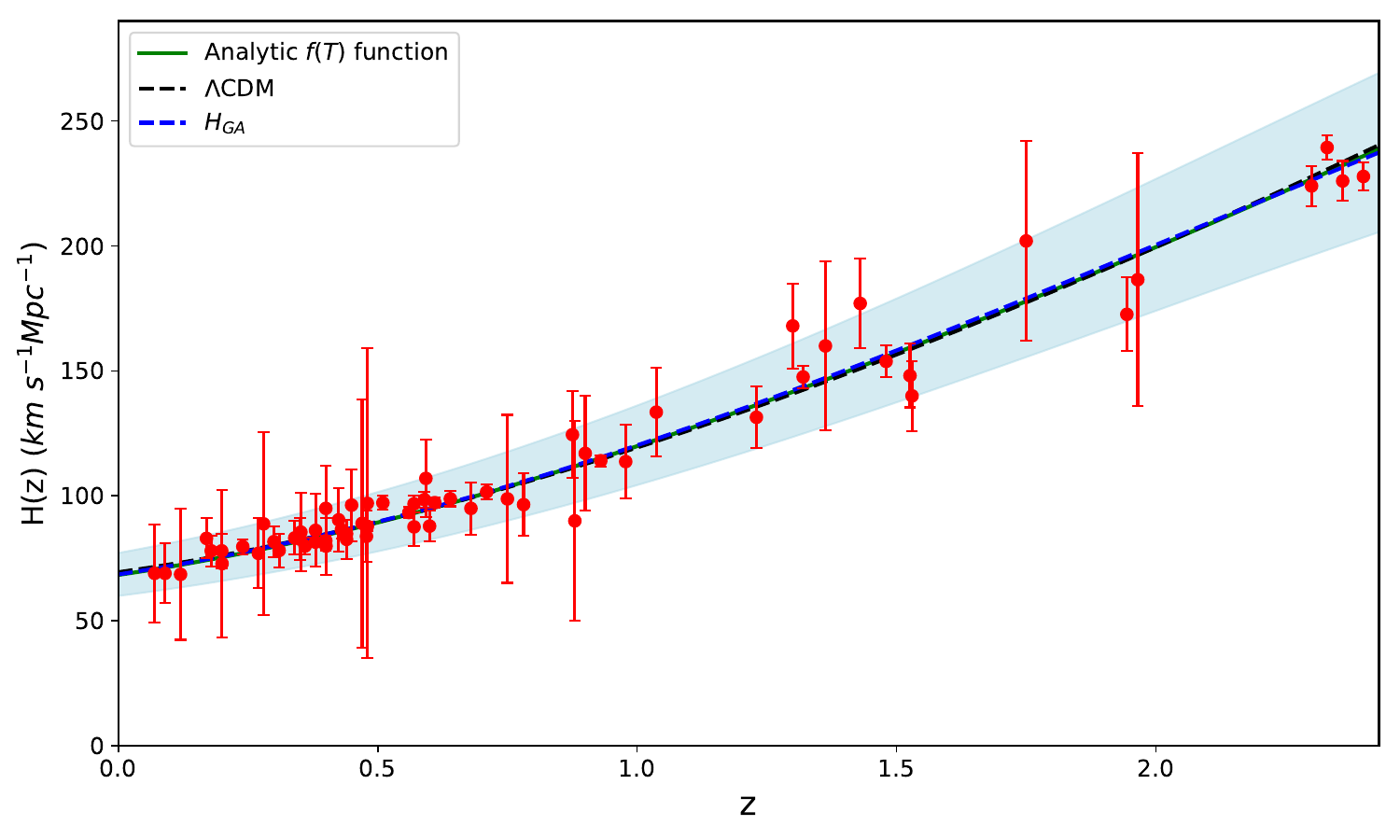}}
		\caption{Left: Independent Reconstruction of the $f(T)$ function from $H(z)$ data using genetic algorithms, imposing $\Omega_{m0}=0.3$. The black curve shows the mean reconstructed curve, the blue area indicates the $1\sigma$ confidence level, the dashed blue curve represents the best-fit analytical model, and the dashed red line corresponds to the $\Lambda$CDM model. Both $T$ and $f(T)$ are divided by $10^5$. Right: Evolution of $H(z)$ from genetic algorithms (dashed blue curve and blue region), the reconstructed $f(T)$ model (green curve), and the $\Lambda$CDM model (dashed black curve).}
		\label{fig}
	\end{figure}

	Our analysis is based on the relationship between the torsion scalar, $T$, of the $f(T)$ gravity and the Hubble parameter, $H$, as described by Eq. (\ref{10}), that is, specifically $T = -6H^2$, which is a simple function of $H$.  Since the dynamics of $f(T)$ are governed by the Friedmann equation, Eq. (\ref{8}), we can express it as a function of the redshift $z$. Consequently, $f_T$ can be expressed as follows
	
	\begin{equation}\label{rr}
		f_T \equiv \frac{df(T)}{dT} = \frac{df/dz}{dT/dz} = \frac{f'}{T'},
	\end{equation}
	wehre the prime denotes derivative with respect to $z$. Substituting Eq. (\ref{rr}) into the Friedmann equation, Eq. (\ref{8}), we can rewrite the latter as
	\begin{equation}\label{20}
		f'(z) = 6 \frac{H'(z)}{H(z)} \left[ H^2(z) - \frac{8 \pi G}{3} \rho_m(z) + \frac{f(z)}{6}  \right].
	\end{equation}

Finally, by incorporating the EoS parameter for the matter sector, we arrive at the final equation, which is expressed as follows  

\begin{equation}\label{A0}
		f'(z) = 6 \frac{H'(z)}{H(z)} \left[ H^2(z) - H_0^2 \Omega_{m0}(1+z)^3 + \frac{f(z)}{6}  \right].
	\end{equation}
	
By using the Hubble parameter derived from GA and its derivative, we solve the provided expression numerically. However, to solve this equation, initial conditions must be imposed. By evaluating the Friedmann equation at $z = 0$, where the $\Lambda$CDM model is assumed to dominate, i.e., $f_{T}(z = 0) \approx 0$, we obtain
	
	\begin{equation}
		f(z = 0) = -6 H_{0}^{2} (1 - \Omega_{m0}).
	\end{equation} 
    
Finally, by utilizing Eq. (\ref{A0}) and the relationship between $T$ and $H$, as well as the reconstruction of $H(z)$, we reconstruct the form of $f(T)$ gravity in a model-independent manner. In the left panel of Fig. \ref{fig}, we present the mean curve for the reconstructed $f(T)$ resulting from our analysis, which constitutes the main result of this study, along with the standard $\Lambda$CDM model represented by a dashed red line. The $\Lambda$CDM model corresponds to a constant value, specifically $f(T) = -2\Lambda = -6H_0^2 (1-\Omega_{m0})$. Furthermore, additional information about the functional form of the reconstructed $f(T)$ gravity can be obtained. By applying the curve fitting method, we find that the reconstructed $f(T)$ curve can be well approximated by a second degree polynomial function, expressed as
\begin{equation}\label{22}
    f(T) = \alpha T_0 + \beta T + \gamma T_0 \left(\frac{T}{T_0}\right)^2,
\end{equation}
with the dimensionless parameters $\alpha = 0.69825$, $\beta = -0.00031$, and $\gamma = 0.00153$, and $T_0$ represents the value of $T$ at the current time, the above analytic function efficiently describes the model-independent reconstructed $f(T)$ form (with an accuracy of $R^{2} \approx 0.9999$). This result is also illustrated in the left panel of Fig. \ref{fig}. Moreover, the cosmological constant lies within the $1\sigma$ region of the reconstructed $f(T)$ model over a significant time range.

In the present work, we proposed the parametrization, Eq. (\ref{22}) for the function $f(T)$, motivated by its ability to accurately reproduce the behavior of the reconstructed function derived from Hubble  data through the GA analysis. The effectiveness of a quadratic form was anticipated, since at late times, when $H \approx H_0$ and thus $T \approx T_0$, a Taylor expansion around $T_0$ can be performed, with the quadratic term naturally emerging as the first significant correction beyond the constant and linear terms. This parametrization is further supported by previous studies in the literature, where quadratic forms of $f(T)$ have been shown to effectively describe cosmological dynamics at late times \cite{Iorio, Nashed, Cai:2016ft, Chen, Cai:2019bdh}.

\begin{figure}[!t] % h: ici, t: haut de la page, b: bas de la page, p: page séparée
	%\centering
	\hspace{-0.6 cm}
		\raisebox{-0.33cm}{\includegraphics[width=0.57\textwidth]{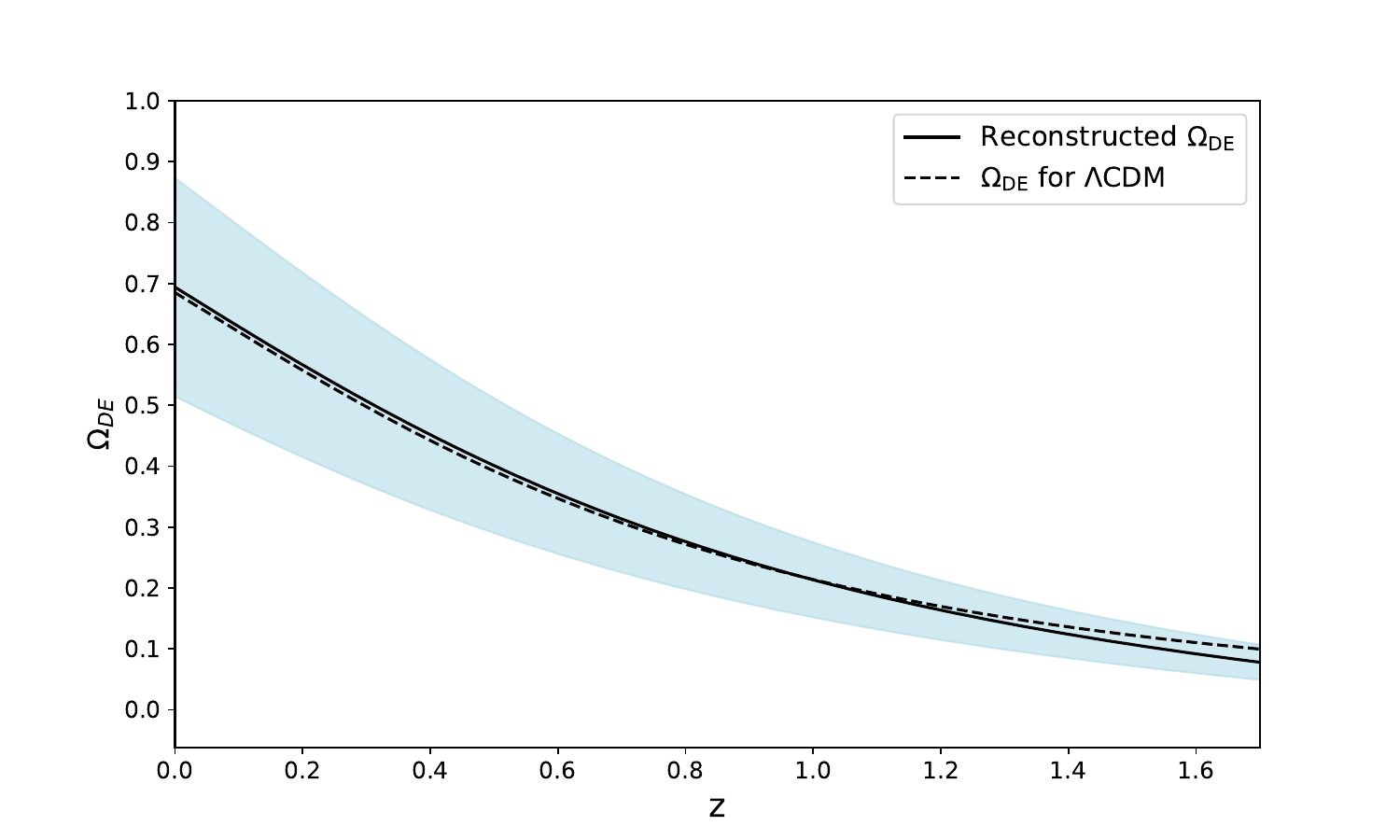}}
		\hspace{-0.7 cm}
		\raisebox{-0.22cm}{\includegraphics[width=0.5\textwidth]{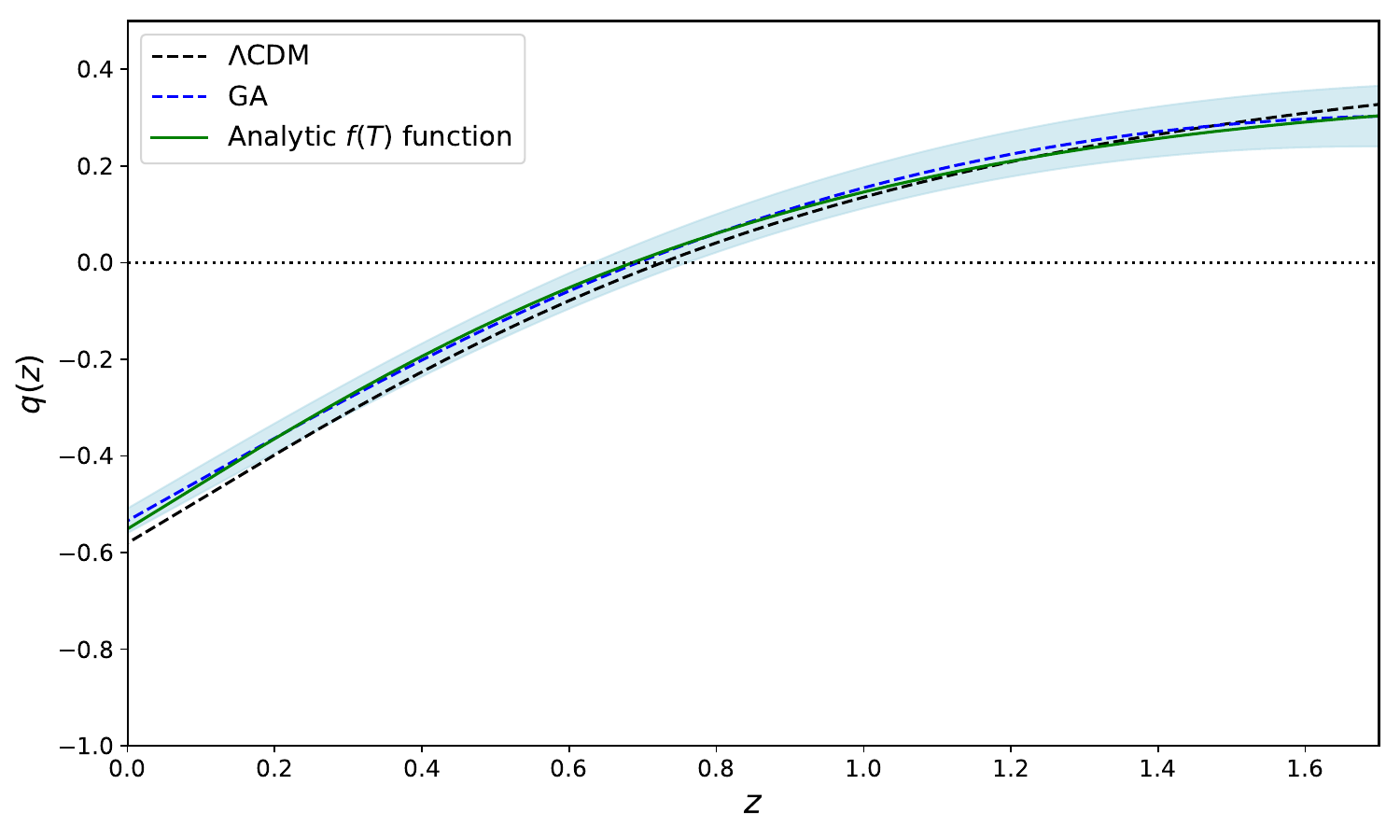}}
	
	\caption{Left: The reconstructed form of the effective dark energy density parameter $\Omega_{DE}$, where the solid black line represents the reconstructed $f(T)$ model, and the blue region corresponds to the $1\sigma$ error. Right: The deceleration parameter $q(z)$, where the green line corresponds to the $f(T)$ model, and the dashed blue line with the blue region show the GA reconstruction and the $1\sigma$ error, respectively. In both panels, the dashed black line represents the $\Lambda$CDM model}

	\label{figgo} 
\end{figure}

As a complementary test of our analysis, we compare the evolution of $H(z)$ derived from the reconstructed $f(T)$ model with that of the $\Lambda$CDM model. For this purpose, we substitute the reconstructed $f(T)$ model into the Friedmann equation, Eq. (\ref{8}), and extract the
solution for $H(z)$. The right panel of Fig. \ref{fig} displays the $H(z)$ curves obtained from the genetic algorithm, the reconstructed $f(T)$ model, and the $\Lambda$CDM model. The results demonstrate strong agreement among the three models. This comparison provides additional validation of our results, further reinforcing the effectiveness of the reconstructed $f(T)$ model in describing the observed data.

To provide further insight into the reconstructed dynamics, we plotted the evolution of the effective dark energy density parameter $\Omega_{DE} = \frac{8\pi G \rho_{DE}}{3 H^2}$ and the deceleration parameter $q(z)$, obtained independently using the reconstruction results of $H(z)$ and $f(T)$. These reconstructions are presented in Fig. \ref{figgo}. In the left panel of Fig. \ref{figgo}, we plot the evolution of $\Omega_{DE}$ as a function of redshift. It is observed that the effective dark energy is currently dominant, with a density parameter of approximately 0.693, which is consistent with current observations. The right panel of Fig. \ref{figgo} illustrates the evolution of the deceleration parameter $q(z)$ for three cases: the $\Lambda$CDM model, the genetic algorithm, and the analytical function $f(T)$ (Eq. \ref{22}). The parameter $q(z)$ is crucial for determining whether the Universe is accelerating and is defined by

	\begin{equation}
		q(z) = -\frac{\ddot{a} a}{\dot{a}^2}\
		= -1 + (1+z) \frac{d \ln H(z)}{dz},
	\end{equation}
	where $a$ represents the scale factor, and the dots indicate derivatives with respect to cosmic time. In the last two scenarios, the deceleration parameter $q(z)$ shifts from positive (indicating deceleration) to negative (indicating acceleration) at low redshifts, similar to what is observed in the $\Lambda$CDM model. This behavior indicates a transition from a decelerating to an accelerating Universe. In both panels of Fig. \ref{figgo}, the blue regions correspond to the  1$\sigma$ confidence level.

\begin{figure}[t]\label{s} 
		\hspace{-0.1cm}
		\raisebox{0.cm}{\includegraphics[width=0.495\textwidth]{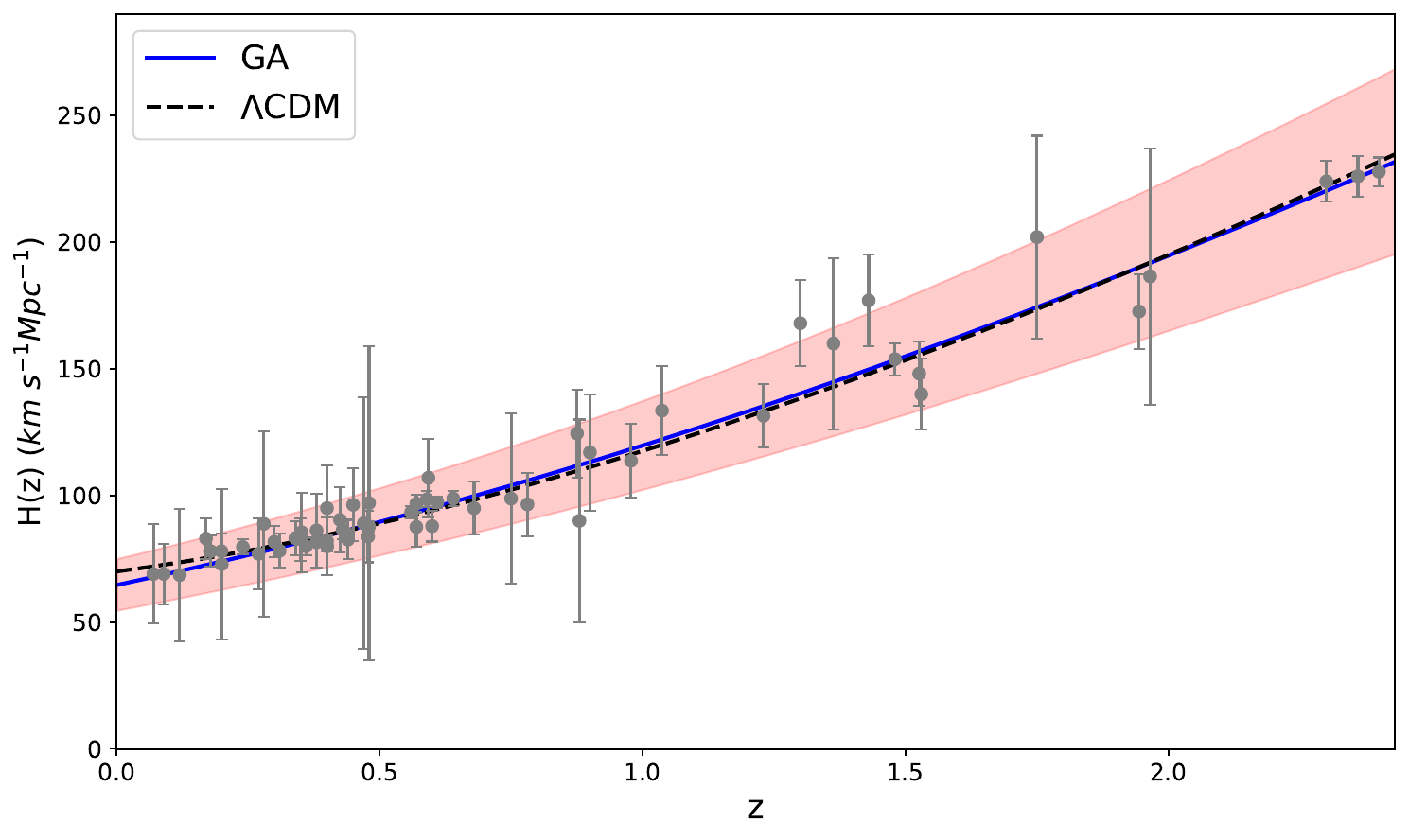}}
		\hspace{-0.cm}
		\raisebox{0. cm}{\includegraphics[width=0.55\textwidth]{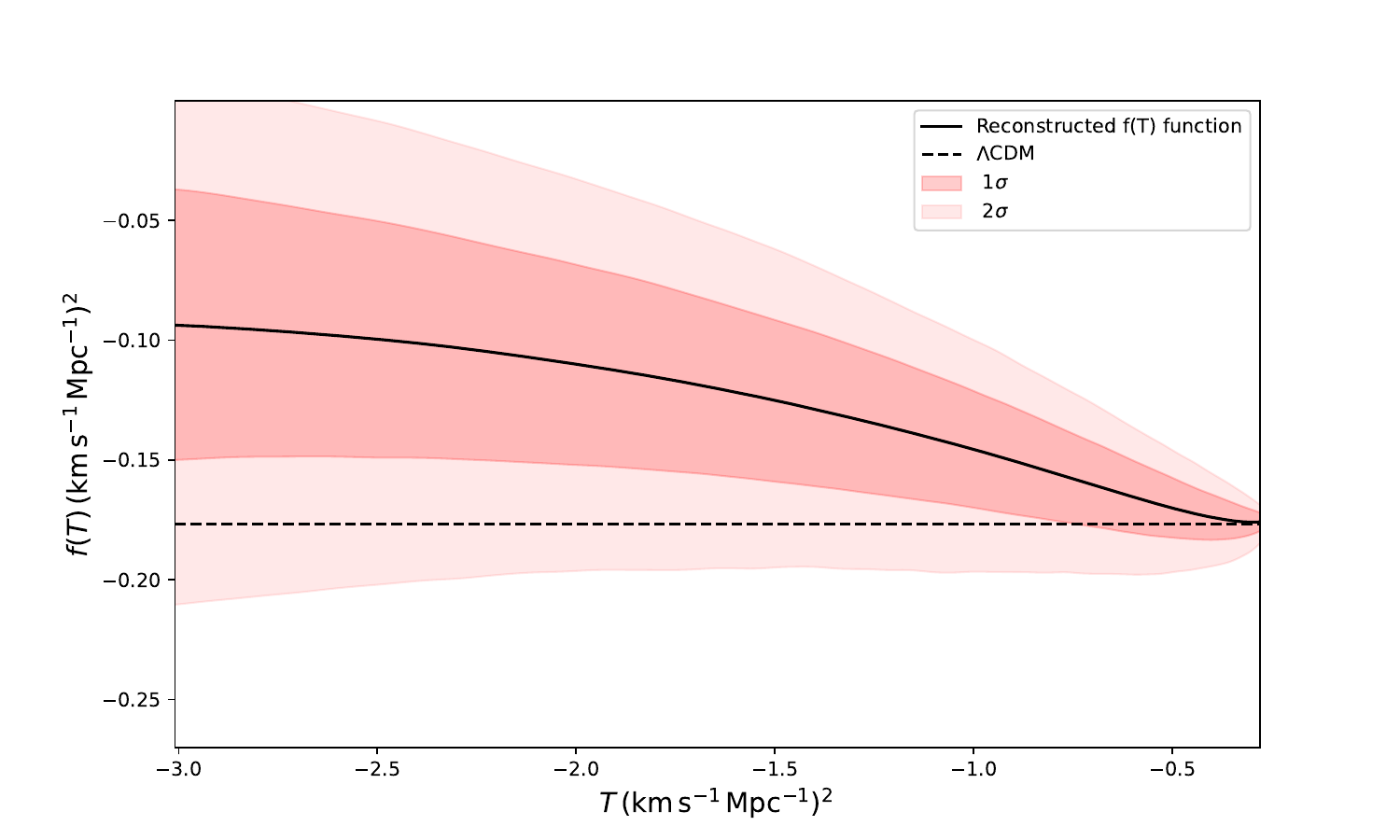}}
		\caption{Left: Compilation of $H(z)$ data excluding DESI measurements, along with the GA best-fit model (blue line). The red region indicates the $1\sigma$ confidence level for the GA model, while the actual data points are depicted as gray points. Right: Independent reconstruction of the $f(T)$ function. The black curve shows the reconstructed function, with light and dark red regions representing the $1\sigma$ and $2\sigma$ confidence levels, respectively. Both $T$ and $f(T)$ are divided by $10^5$. In both panels, the dashed black line represents the $\Lambda$CDM model.}
		\label{fig6}
	\end{figure}

To conclude this subsection, we quantify the impact of DESI data on the reconstruction process by performing a complementary analysis that excludes DESI measurements. The Hubble parameter, $H(z)$, reconstructed from GA, is presented in the left panel of Fig.~\ref{fig6}, yielding $H_0 = 64.62 \pm 9.98 \, \text{km/s/Mpc}$. This relatively low value, accompanied by a large uncertainty, is primarily driven by the  previous BAO measurements, which are known to favor lower values of $H_0$~\cite{Yang:2024kdo}. In contrast, incorporating DESI data, particularly its high-precision measurements at intermediate and high redshifts, yields a more precise reconstruction, with $H_0 = 68.56 \pm 8.59 \, \text{km/s/Mpc}$, as discussed in Subsection~\ref{subsection3.3}. This result aligns more closely with local determinations and demonstrates a 13.9\% reduction in uncertainty, highlighting the enhanced constraining power provided by DESI. A comparable behavior is observed in the reconstruction of the $f(T)$ function. As shown in the right panel of Fig.~\ref{fig6}, excluding DESI data leads to larger deviations from the $\Lambda$CDM model and broader confidence intervals, especially at high redshifts, where observational constraints are otherwise limited. In contrast, the reconstruction obtained with DESI data, as shown in the left panel of Fig.~\ref{fig}, remains well within the $1\sigma$ region around $\Lambda$CDM model across a wide redshift range, indicating improved stability and accuracy. These findings demonstrate the crucial role of DESI in enhancing the robustness, precision, and fidelity of the reconstruction.

	\section{Conclusion}\label{section5}\
	
	In this work, we used data from the Hubble parameter $H(z)$ and genetic algorithm to reconstruct the function $f(T)$ in a model-independent manner. GA analysis was applied to Hubble measurements, including Cosmic Chronometers, Baryon Acoustic Oscillations, and the latest cosmological data released by DESI collaboration, which allowed us to reconstruct $H(z)$ without relying on a specific model as presented in Section \ref{section3}. We derived a Hubble parameter function that more accurately describes the observational data. Moreover, $f(T)$ gravity theory has the advantage that the torsion scalar is a simple function of $H(z)$, namely $T = -6H^2$. Therefore, knowing $H(z)$ and the value of the Hubble constant derived by genetic algorithm, $H_0 = (68.56 \pm 8.59)$ km/s/Mpc, we  reconstructed $f(T)$, which is a key result of this work.

	The left panel of Fig. \ref{fig} illustrates the reconstructed function $f(T)$ as a function of the torsion scalar, $T$, where the black curve represents the mean reconstructed function, and the blue region indicates the $1\sigma$ confidence level. The red dashed line represents the $\Lambda$CDM model, which is a constant, i.e., $f_{\Lambda\text{CDM}}(T) = -2\Lambda = -6 H_{0}^{2}(1-\Omega_{m0})$. Additionally, the functional form of the reconstructed $f(T)$ gravity provides further insight. Using a curve fitting method, we find that the reconstructed $f(T)$ curve is well approximated by a second degree polynomial function.

	Afterwards, we used the reconstructed  $f(T)$ model to compare the evolution of the parameter $H(z)$ with that of the $\Lambda$CDM using the Friedmann equation. The right panel of Fig. \ref{fig} presents the $H(z)$ curves obtained from the genetic algorithm, the reconstructed $f(T)$ model, and the $\Lambda$CDM model. We observe significant agreement among the three models. This comparison serves as additional cross validation of our results, further reinforcing the effectiveness of the $f(T)$ model in describing the observational data. We also used the reconstructed $f(T)$ model to plot the effective dark energy density parameter, and the deceleration parameter. We observed that, these quantities align with the predictions of the  $\Lambda$CDM model. The reconstruction of these quantities as a function of redshift $z$, along with the $1\sigma$ confidence level, is presented in Fig. \ref{figgo}.\

Furthermore, to assess the impact of the DESI data on the reconstruction, we performed a complementary analysis excluding these measurements. We found that the inclusion of DESI data significantly improves the precision and robustness of the reconstruction.

\color{black}

	In summary, we followed a reconstruction procedure that is completely model-independent, based solely on Hubble data, including Cosmic Chronometers, Baryon Acoustic Oscillations, and the latest cosmological data released by DESI collaboration. This paper presents the first implementation of GA to reconstruct the form of $f(T)$ gravity. Moreover, our method offers several advantages over traditional or non-parametric models such as cosmography, which faces convergence issues at high redshifts, or Gaussian processes, which require the choice of a kernel function and a reference model \cite{arjona2020can}. Our approach neither requires a kernel function nor a reference model and remains entirely model-independent. We demonstrated the effectiveness of GA in reconstructing the function $f(T)$ in a model-independent way. The reconstructed $f(T)$ model is in excellent agreement with the observed data. Our perspective for upcoming work is to focus on extending our GA analysis to include additional datasets, such as supernovae and large-scale structure data. This will allow us to obtain more precise results and open new avenues for exploring solutions to unresolved problems in cosmology, such as the Hubble tension $H_0$ and $\sigma_8$, within the framework of modified gravity theories.

	%\newpage
	\section*{Acknowledgements}\
    
	R. E. O. would like to thank Savvas Nesseris for his insightful discussions and valuable suggestions regarding genetic algorithms, and also Xin Ren and Emmanuel N. Saridakis for their helpful discussions concerning  $f(T)$ gravity. R. E. O. is grateful for support from the \enquote{PhD-Associate Scholarship – PASS} grant (number 29 UMP2023) of the National Center for Scientific and Technical Research in Morocco.


\begin{thebibliography}{10}
		
		\bibitem{SupernovaSearchTeam:1998fmf}
A.~G. Riess et al., Astron. J. \textbf{116}, 1009--1038 (1998).

\bibitem{SupernovaCosmologyProject:1998vns}
S. Perlmutter et al., Astrophys. J. \textbf{517}, 565--586 (1999).

\bibitem{Wong:2019kwg}
K.~C. Wong et al., Mon. Not. R. Astron. Soc. \textbf{498}, 1420--1439 (2020).

\bibitem{Planck:2018vyg}
N. Aghanim et al., Astron. Astrophys. \textbf{641}, A6 (2020).

\bibitem{Riess:2021jrx}
A.~G. Riess et al., Astrophys. J. Lett. \textbf{934}, L7 (2022).

\bibitem{Poulin:2022sgp}
V. Poulin et al., Phys. Rev. D \textbf{107}, 123538 (2023).

\bibitem{Dahmani:2023bsb}
S.~Dahmani, A.~Bouali, I.~El Bojaddaini et al., Phys. Dark Univ. \textbf{42}, 101266 (2023).


		
	\bibitem{Abdalla:2022yfr}
E.~Abdalla, G.~Franco Abell\'an, A.~Aboubrahim et al., JHEAp \textbf{34}, 49--211 (2022).

\bibitem{mhamdi2023comparing}
D.~Mhamdi, F.~Bargach, S.~Dahmani et al., Gen. Rel. Grav. \textbf{55}, 11 (2023).

\bibitem{Karwal:2016vyq}
T. Karwal and M. Kamionkowski, Phys. Rev. D \textbf{94}, 103523 (2016).

\bibitem{Vagnozzi:2021gjh}
S. Vagnozzi, Phys. Rev. D \textbf{104}, 063524 (2021).

\bibitem{bouhmadi2018more}
M. Bouhmadi-López,  A. Errahmani, T. Ouali et al., Eur. Phys. J. C \textbf{78}, 1--11 (2018).

		
		\bibitem{errahmani2006high}
A. Errahmani and T. Ouali, Phys. Lett. B \textbf{641}, 357--361 (2006).

\bibitem{enkhili2024diagnostic}
O. Enkhili, F.~Bargach, D.~Mhamdi et al., New Astronomy \textbf{113}, 102298 (2024).

\bibitem{belkacemi2020interacting}
M. H. Belkacemi, Z.~Bouabdallaoui, M.~Bouhmadi-L\'opez et al., Int. J. Mod. Phys. D \textbf{29}, 2050066 (2020).

\bibitem{Gelmini:2019deq}
G. B. Gelmini, A. Kusenko, and V. Takhistov, JCAP \textbf{06}, 002 (2021).

\bibitem{Errahmani:2024ran}
A. Errahmani,  A.~Bouali, S.~Dahmani et al., Phys. Dark Univ. \textbf{45}, 101512 (2024).

\bibitem{mhamdi2024cosmological}
D. Mhamdi, A.~Bouali, S.~Dahmani et al., Eur. Phys. J. C \textbf{84}, 310 (2024).

		
		\bibitem{bouali2023observational}
A. Bouali, H.~Chaudhary, T.~Harko et al., Mon. Not. Roy. Astron. Soc. \textbf{526}, 4192--4208 (2023).

\bibitem{Koussour:2022dcb}
M. Koussour, S.~H.~Shekh, M.~Bennai et al., Chin. J. Phys. \textbf{90}, 97--107 (2024).

\bibitem{harko2022palatini}
T. Harko, F. S. N. Lobo, S. Nojiri et al., Phys. Rev. D \textbf{84}, 024020 (2011).

\bibitem{Clifton:2011jh}
T. Clifton, P.~G.~Ferreira, A.~Padilla et al., Phys. Rept. \textbf{513}, 1--189 (2012).

\bibitem{bargach2021dynamical}
F. Bargach, A.~Bargach, M.~Bouhmadi-L\'opez et al., Int. J. Mod. Phys. D \textbf{30}, 2150076 (2021).

\bibitem{ElOuardi}
R. El Ouardi, A.~Bouali, S.~Dahmani et al., Phys. Lett. B \textbf{863}, 139374 (2025).


		
		\bibitem{DeFelice:2010aj}
A. De Felice and S. Tsujikawa, Living Rev. Rel. \textbf{13}, 3 (2010).

\bibitem{Nojiri:2005jg}
S. Nojiri and S. D. Odintsov, Phys. Lett. B \textbf{631}, 1--6 (2005).

\bibitem{ladghami20234d}
Y. Ladghami, B.~Asfour, A.~Bouali et al., Phys. Dark Univ. \textbf{41}, 101261 (2023).



\bibitem{Maluf:2013gaa}
J. W. Maluf, Annalen Phys. \textbf{525}, 339--357 (2013).

\bibitem{Aldrovandi:2013wha}
R. Aldrovandi and J. G. Pereira, Teleparallel Gravity: An Introduction (Dordrecht: Springer, 2013).

		
		\bibitem{Krssak:2018ywd}
M. Krssa, R.~J.~van den Hoogen, J.~G.~Pereirak et al., Class. Quant. Grav. \textbf{36}, 183001 (2019).

\bibitem{Bengochea:2008gz}
G. R. Bengochea and R. Ferraro, Phys. Rev. D \textbf{79}, 124019 (2009).

\bibitem{Linder:2010py}
E. V. Linder, Phys. Rev. D \textbf{81}, 127301 (2010).

\bibitem{Kofinas:2014owa}
G. Kofinas and E. N. Saridakis, Phys. Rev. D \textbf{90}, 084044 (2014).



\bibitem{Capozziello:2016eaz}
S. Capozziello, M.~De Laurentis and K.~F.~Dialektopoulos et al., Eur. Phys. J. C \textbf{76}, 629 (2016).

\bibitem{Heisenberg:2023lru}
L. Heisenberg, Phys. Rept. \textbf{1066}, 1--78 (2024).

\bibitem{enkhili2024cosmological}
O. Enkhili, S. Dahmani, D. Mhamdi et al., Eur. Phys. J. C \textbf{84}, 1--11 (2024).

\bibitem{koussour2023cosmic}
M. Koussour, S.~Dahmani, M.~Bennai et al., Eur. Phys. J. Plus \textbf{138}, 179 (2023).

\bibitem{mhamdi2024observational}
D. Mhamdi, S.~Dahmani, A.~Bouali et al., arXiv preprint arXiv:2408.11996 (2024).

\bibitem{Ferraro:2006jd}
R. Ferraro and F. Fiorini, Phys. Rev. D \textbf{75}, 084031 (2007).

		
		\bibitem{Ferraro:2008ey}
R. Ferraro and F. Fiorini, Phys. Rev. D \textbf{78}, 124019 (2008).

\bibitem{Cai:2019bdh}
Y.-F. Cai, M.~Khurshudyan and E.~N.~Saridakis, Astrophys. J. \textbf{888}, 62 (2020).

\bibitem{Nesseris:2013jea}
S. Nesseris, S.~Basilakos, E.~N.~Saridakis et al., Phys. Rev. D \textbf{88}, 103010 (2013).

\bibitem{moriwaki2023machine0}
K. Moriwaki, T.~Nishimichi and N.~Yoshida, Rep. Prog. Phys. \textbf{86}, 076901 (2023).

\bibitem{Peel:2018aei}
A. Peel, F.~Lalande, J.~L.~Starck et al., Phys. Rev. D \textbf{100}, 023508 (2019).

\bibitem{Ho:2019zap}
M. Ho, M. M. Rau, M. Ntampaka et al., Astrophys. J. \textbf{887}, 25 (2019).

\bibitem{Akrami:2009hp}
Y. Akrami,  P. Scott, J. Edsjö et al., JHEP \textbf{04}, 057 (2010).

\bibitem{Allanach:2004my}
B. C. Allanach, D. Grellscheid and F. Quevedo, JHEP \textbf{07}, 069 (2004).

		
		\bibitem{Nesseris:2010ep}
S. Nesseris and A. Shafieloo, Mon. Not. Roy. Astron. Soc. \textbf{408}, 1879 (2010).



\bibitem{Nesseris:2013bia}
S. Nesseris and J. García-Bellido, Phys. Rev. D \textbf{88}, 063521 (2013).


\bibitem{Alestas:2022ml}
G. Alestas, L. Kazantzidis and S. Nesseris, Phys. Rev. D \textbf{106}, 103519 (2022).



\bibitem{rajpaul2012genetic}
P. Charbonneau, Astrophys. J. Supp. \textbf{101}, 309 (1995).

\bibitem{Ren:2022aeo}
X. Ren, S.-F. Yan, Y. Zhao et al., Astrophys. J. \textbf{932}, 2 (2022).

\bibitem{Ren:2021tfi}
X. Ren, T. H. T. Wong, Y.-F. Cai et al., Phys. Dark Univ. \textbf{32}, 100812 (2021).

\bibitem{fortunato2024search}
J. A. S. Fortunato, P.~H.~R.~S.~Moraes, J.~G.~d.~J\'unior et al., Eur. Phys. J. C \textbf{84}, 198 (2024).

\bibitem{Arjona:2020kco}
R. Arjona and S. Nesseris, JCAP \textbf{11}, 042 (2020).

\bibitem{Bernardo} R.C. Bernardo and J.L. Said, J. Cosm. Astropart. Phys. \textbf{2021}, 027 (2021).




\bibitem{gaztanaga2009clustering} E. Gaztanaga, A.~Cabre and L.~Hui, Mon. Not. Roy. Astron. Soc. \textbf{399}, 1663--1680 (2009).

		
\bibitem{Jimenez:2001gg}
R. Jimenez and A. Loeb, Astrophys. J. \textbf{573}, 37 (2002).

\bibitem{adame2024desi}
A. G. Adame et al., JCAP \textbf{2025}, 021 (2025).


\bibitem{t}
R. Weitzenböck, Invarianten Theorie, Nordhoff, Groningen (1923).


%%%%%%%%%%%%%%%%%%%%%%%%%%%%%% Table
\bibitem{zhang2014four1} C. Zhang, H.~Zhang, S.~Yuan et al., Res. Astron. Astrophys. \textbf{14}, 1221 (2014).



\bibitem{jimenez2003constraints} R. Jimenez, L.~Verde, T.~Treu et al., Astrophys. J. \textbf{593}, 622 (2003).


\bibitem{stern2010cosmic} J. Simon, L.~Verde and R.~Jimenez, Phys. Rev. D \textbf{71}, 123001 (2005).







%%%%%%%%%%%%%%%
\bibitem{moresco2012improved} M. Moresco, A. Cimatti, R. Jimenez et al., JCAP \textbf{1208}, 006 (2012).

\bibitem{Moresco2016mzx} M. Moresco, L. Pozzetti, A. Cimatti et al., JCAP \textbf{1605}, 014 (2016).

\bibitem{ratsimbazafy2017age}
A. L. Ratsimbazafy, S.~I.~Loubser, S.~M.~Crawford et al., Mon. Not. Roy. Astron. Soc. \textbf{467}, 3239--3254 (2017). 

\bibitem{stern2010cosmic1} D. Stern, R. Jimenez, L. Verde et al., JCAP \textbf{2}, 8 (2010).


\bibitem{borghi2022toward}
N. Borghi, M.~Moresco and A.~Cimatti, Astrophys. J. Lett. \textbf{928}, L4 (2022).





\bibitem{Moresco2015cya} M. Moresco, Mon. Not. Roy. Astron. Soc. \textbf{450}, L16 (2015).

\bibitem{oka2014simultaneous}
A. Oka, S.~Saito, T.~Nishimichi, Mon. Not. Roy. Astron. Soc. \textbf{439}, 2515--2530 (2014).








\bibitem{wang2017clustering}
Y. Wang et al., Mon. Not. Roy. Astron. Soc. \textbf{469}, 3762--3774 (2017).

\bibitem{chuang2013modelling}
C.-H. Chuang and Y.~Wang, Mon. Not. Roy. Astron. Soc. \textbf{435}, 255--262 (2013).

\bibitem{alam2017clustering}
S. Alam et al., Mon. Not. Roy. Astron. Soc. \textbf{470}, 2617–2652 (2017).

\bibitem{blake2012wigglez}
C. Blake, S.~Brough, M.~Colless et al., Mon. Not. Roy. Astron. Soc. \textbf{425}, 405--414 (2012).

\bibitem{chuang2013clustering}
C.-H. Chuang, F.~Prada, A.~J.~Cuesta et al., Mon. Not. Roy. Astron. Soc. \textbf{433}, 3559--3571 (2013).













\bibitem{anderson2014clustering}
L. Anderson et al., Mon. Not. Roy. Astron. Soc. \textbf{441}, 24--62 (2014).

\bibitem{zhao2019clustering}
G. B. Zhao et al., Mon. Not. Roy. Astron. Soc. \textbf{482}, 3497--3513 (2019).

\bibitem{neveux2020completed}
R. Neveux et al., Mon. Not. Roy. Astron. Soc. \textbf{499}, 210–229 (2020).




\bibitem{Busca_2013}
N. G. Busca, T. Delubac, J. Rich et al., Astron. Astrophys. \textbf{552}, A96 (2013).


\bibitem{font2014quasar}
A. Font-Ribera et al., J. Cosm. Astropart. Phys. \textbf{2014}, 027 (2014).

\bibitem{des2017baryon}
H. D. Mas Des Bourboux, J.-M. Le Goff, M. Blomqvist et al., Astron. Astrophys. \textbf{608}, A130 (2017).
%%%%%%%%%%%%%%%%%%%%%%%%%%%%%%%%%%%%%%%%%%%%%%%%%%%%%%%%%%%%%%%%%%%%%%%%%%%%%%%%%%%%%%%%%%%%%%%%%%%


\bibitem{bogdanos2009genetic}
C. Bogdanos and S. Nesseris, JCAP \textbf{05}, 006 (2009).

\bibitem{Nesseris_2012}
S. Nesseris and J. García-Bellido, JCAP \textbf{2012}, 033 (2012).


\bibitem{Yang:2024kdo}
Y. Yang, X.~Ren, Q.~Wang et al., Sci. Bull. \textbf{69}, 2698 (2024).







\bibitem{gil2020completed}
H. Gil-Marín et al., Mon. Not. Roy. Astron. Soc. \textbf{498}, 2492–2531 (2020).

\bibitem{bautista2021completed}
J. E. Bautista et al., Mon. Not. Roy. Astron. Soc. \textbf{500}, 736–762 (2021).

\bibitem{avila2020completed}
S. Avila et al., Mon. Not. Roy. Astron. Soc. \textbf{499}, 5486–5507 (2020).

\bibitem{hou2021completed}
J. Hou et al., Mon. Not. Roy. Astron. Soc. \textbf{500}, 1201–1221 (2021).



\bibitem{des2020completed}
H. Du Mas Des Bourboux et al., Astrophys. J. \textbf{901}, 153 (2020).

\bibitem{moresco2020setting}
M. Moresco, R.~Jimenez, L.~Verde et al., Astrophys. J. \textbf{898}, 82 (2020).


\bibitem{vazirnia}
M. Vazirnia and A. Mehrabi, Phys. Rev. D \textbf{104}, 123530 (2021).

\bibitem{mukherjee}
P. Mukherjee and A. Mukherjee, Mon. Not. R. Astron. Soc. \textbf{504}, 3938 (2021).


\bibitem{Iorio}
L. Iorio and E. N. Saridakis, Mon. Not. Roy. Astron. Soc. \textbf{427}, 1555--1561 (2012).

\bibitem{Nashed}
G. L. Nashed, Gen. Rel. Grav. \textbf{47}, 75 (2015).

\bibitem{Cai:2016ft}
Y.-F. Cai, S.~Capozziello, M.~De Laurentis et al., Rept. Prog. Phys. \textbf{79}, 106901 (2016).



\bibitem{Chen}
Z. Chen, W.~Luo, Y.~F.~Cai et al., Phys. Rev. D \textbf{102}, 104044 (2020).











        













%%%%%%%%%%%%%%%%%%%%%%%%%

\bibitem{arjona2020can}
R. Arjona and S. Nesseris, Phys. Rev. D \textbf{101}, 123525 (2020).

	\end{thebibliography}
\end{document}